\documentclass[prb,superscriptaddress,twocolumn,amsmath,amssymb,floatfix]{revtex4-1}
\usepackage{latexsym}
\usepackage{dcolumn}
\usepackage[dvips]{graphicx}
\usepackage{amssymb}
\usepackage{graphics}
\usepackage{amsmath}
\usepackage{epsf}

\begin{document}
\title{Valley-dependent 2D transport in Si-MOSFETs}
\author{E. H. Hwang}
\affiliation{Condensed Matter Theory Center, Department of Physics, 
	 University of Maryland,
	 College Park, Maryland  20742-4111}
\affiliation{SKKU Advanced Institute of Nanotechnology, Sungkyunkwan
  University, Suwon 440-746, Korea } 
\author{S. Das Sarma}
\affiliation{Condensed Matter Theory Center, Department of Physics, 
	 University of Maryland,
	 College Park, Maryland  20742-4111}

\date{\today}

\begin{abstract}
Motivated by interesting recent experimental results, we consider
theoretically charged-impurity scattering-limited 2D electronic
transport in (100), (110), and (111)-Si inversion layers at low temperatures
and carrier densities, where screening effects are important. We show
conclusively that, given the same bare Coulomb disorder, the 2D
mobility for a given system increases monotonically with
increasing valley degeneracy. We also show that the temperature and
the parallel magnetic field dependence of the 2D conductivity is
strongly enhanced by increasing valley degeneracy. We analytically
consider the low temperature limit of 2D
transport, particularly its theoretical dependence on valley
degeneracy, comparing with our full numerical results and with the
available experimental results. We make qualitative and quantitative
predictions for the parallel magnetic field induced 2D
magnetoresistance in recently fabricated high-mobility 6-valley
Si(111)-on-vacuum inversion layers. We also provide a theory for 2D
transport in ultrahigh mobility Si(111) structures recently fabricated
in the laboratory, discussing the possibility of observing the
fractional quantum Hall effect in such Si(111) structures.
\end{abstract}

\maketitle

\section{introduction}

Many semiconductor-based two dimensional (2D) electron systems have an
intrinsic valley degeneracy ($g_v$) in addition to the spin degeneracy
($g_s=2$). This valley degeneracy, which arises from the bulk band
structure of the corresponding 3D material, is usually exact within
the effective mass approximation, but is only approximate in the
experimental 2D systems where there could be small energy level
splittings between different valleys (the so-called ``valley
splitting'')\cite{ando}. If the valley splitting is ``small'' in some operational
sense, the 2D system could be considered to have a total quantum
degeneracy of $g=g_sg_v$ with both spin and valley states being
quantum degenerate. A well known example of valley degeneracy
is graphene, which has a valley degeneracy of $g_v=2$ with equivalent
Dirac cones at $K$ and $K'$ points of the Brillouin
zone\cite{dassarmarmp}. Examples of 
semiconductor-based 2D electron systems with valley degeneracy
$g_v(>1)$ are Si(100), (110), and (111) electron inversion layers in
MOSFETs\cite{ando} as well as AlAs- and AlSb-based 2D electron
systems\cite{vurgaftman,Kesteren}. Many 
other 2D systems (e.g. n-GaAs and p-GaAs 2D electron and hole systems)
have no valley degeneracy ($g_v=1$)\cite{vurgaftman}. Valley
degeneracy obviously has a 
profound effect on the electronic properties of the 2D system. The
purpose of the current paper is a systematic theoretical investigation of the
valley degeneracy effect on 2D electronic transport properties, using
Si-MOSFET structure based n-inversion layers as the specific system
under consideration (since these typically have $g_v>1$) although our
qualitative and analytical results would apply to all 2D semiconductor
systems (specifically Si-Ge 2D electron systems) with $g_v \geq 1$.

The valley-dependent electronic properties of a 2D (or 3D for that
matter) semiconductor system is best understood by considering the
electronic density of states in the valley-degenerate ground state,
which is given for the 2D (3D) system by: $D(\epsilon)=g m/2\pi
\hbar^2$ ($g\sqrt{\epsilon}(2m)^{3/2}/4\pi^2\hbar^3$), where
$g=g_sg_v$ is the ground state degeneracy arising from both spin
($g_s$) and valley ($g_v$) degeneracies. The linear proportionality of
the electronic density of states with the valley degeneracy leads
immediately to the following dependences of the Fermi wave vector
($k_F$), the Thomas-Fermi screening wave vector ($q_{TF}$), and the
Fermi energy ($E_F$) on the valley degeneracy in 2D and 3D systems:
\begin{subequations}
\begin{eqnarray}
k_F & \sim & g_v^{-1/2} \;\; {\rm (2D)};\;\;\;   g_v^{-1/3} \;\; {\rm
  (3D)}, \\
q_{TF} & \sim & g_v \;\;\;\;\;\;\; {\rm (2D)};\;\;\;   g_v^{2/3}
\;\;\;\; {\rm (3D)}, \\
\epsilon_F & \sim & g_v^{-1} \;\;\;\;\; {\rm (2D)};\;\;\;  
g_v^{-2/3} \;\; {\rm (3D)}.
\end{eqnarray}
\end{subequations}
Since $D(\epsilon)$, $k_F$, $q_{TF}$, and $\epsilon_F$ all depend
nontrivially on the valley degeneracy factor $g_v$, all electronic
properties, including 2D transport properties, depend nontrivially on
the valley degeneracy. We note that the valley-dependent transport
properties are in general nontrivial since $g_v$ enters independently
through both $q_{TF}$ (and hence screening) and $k_F$ (and hence
scattering wave vector). Increasing (decreasing) the valley degeneracy
enhances (suppresses) screening through $q_{TF}$, but at the same time
it also affects $k_F$, increasing it with decreasing $g_v$.

Bulk Si has six equivalent conduction band minima located about 85\%
to the Brillouin zone boundary, thus making bulk Si a $g_v=6$
system. Each valley corresponds to an ellipsoid with anisotropic
effective mass along and perpendicular to the symmetry axes. There
are thus three possible Si-based 2D electron systems, depending on
whether (100), (110), or (111) surface is used for creating the 2D
confinement. Within the effective mass approximation these three
distinct 2D Si systems have different valley degeneracies: $g_v=2$
[Si(100)], 4 [Si(110)], and 6 [Si(111)]. Uniaxial stress would lift the
4-fold or the 6-fold valley degeneracy of Si(110) or (111) system, making
ground state of each a doubly degenerate $g_v=2$ system similar to the
Si(100) system (but, of course, with a distinct effective mass for each 2D
system). It is important to mention in this context the fact that
essentially all Si-based 2D systems studied in the literature,
independent of the surface orientation, have almost universally always
experimentally manifested $g_v=2$, even for Si(110) and (111) systems, which
nominally should have $g_v=4$ and 6, respectively. This is thought to
be due to extensive random uniaxial stress universally present at the
Si-SiO$_2$ interface in Si MOSFET structures, which pulls down two
equivalent valleys compared with the other valleys, making both Si(110)
and Si(111) MOSFETs to have doubly degenerate ground states ($g_v=2$)
similar to the Si(100) MOSFETs (but with different effective
masses). An early experiment \cite{kaminsky} did manage to observe a
six-fold valley 
degeneracy in Si(111)- SiO$_2$ MOSFETs, but the system had very poor
mobility and was not useful for the investigation of valley-dependent
transport properties. Other than this one exception \cite{kaminsky},
all Si-SiO$_2$ 2D 
MOSFETs invariably manifest $g_v=2$ ground state, independent of their
surface orientation in sharp contrast to the effective mass
approximation based expectation of $g_v=6$ (4) for Si(111) [(110)]
systems.

An exciting new experimental development in the subject, which is the
direct motivation for our study, is the recent fabrication of very
high quality Si(111)-on-vacuum 2DEG FET structures \cite{Eng,Eng1},
which exhibit 
$g_v=6$ ground state in agreement with the Si bulk band structure
effective mass approximation. Presumably the very high quality
(without any interface strain) of the Si-vacuum interface, leading to
very high mobility ($\sim 10^5$ cm$^2$/Vs), produce the expected 6-fold
valley degeneracy. The fact that these Si-vacuum 2D FET systems
also have very high mobility is consistent with the high quality of
the Si surface leading to the $g_v=6$ Si(111) 2D system. The
absence of a Si-SiO$_2$ interface may be the reason that these new
Si-vacuum based 2D systems satisfy the expected $g_v=6$ effective mass
approximation prediction. 
The absence of a real solid interface may simply enable the bulk
effective mass approximation to be valid at the surface leading to the
Si(111) 6-fold degeneracy.
Although similar 2D systems on
Si(110)-vacuum system have not yet been made, it is reasonable to
expect that the corresponding Si(110) 2D system will have $g_v=4$ valley
degeneracy. We note that these Si-vacuum 2D systems manifest, in
addition to the expected valley degeneracy anticipated on the basis of
the effective mass approximation, also extremely high 2D mobilities
because of the lack of random impurities in the oxide layer which
adversely affects the mobility in the Si-SiO$_2$ 2D systems.

In particular, of course, the effective mass approximation is not
exact and  breaks down at an interface\cite{sham}. Thus, even the double
degeneracy ($g_v=2$) of the Si(100) 2D system [or of the Si(110) and (111)
systems as observed experimentally] is only approximate and is lifted
beyond the simple effective mass approximation leading to small ($<
1$meV) energy splitting between the two valleys. One can, therefore, 
think about an experimental single valley 2D Si system where this ground
state valley splitting is large enough so
that the higher valley state is not occupied by electrons.
Thus, in principle, the valley degeneracy of Si-based 2D MOSFET or
inversion layer system for any surface orientation can be thought
to be a continuous variable ranging between 1 and 6 depending
on the microscopic details of the interface. This is the approach we
take in the current work where $g_v$ is assumed to be a free
parameters to be determined experimentally.

Motivated by the above considerations, we theoretically consider
valley dependent 2D transport in Si systems assuming the valley
degeneracy $g_v$ to be a free rational variable --- in reality, of
course, $g_v$ can only be 1, 2, 4, or 6 in Si 2D systems depending on
the situation. (In some cases, where the valleys are occupied by
unequal number of electrons because of small valley splittings, it may
be useful to think of the valley degeneracy being a fractional
number.) We address the density, the temperature, and the in-plane
magnetic field dependence of 2D transport in the presence of a variable
valley degeneracy. The applied in-plane magnetic field is parallel to
the 2D system and is therefore assumed to only affect the spin
degeneracy of the 2DEG since it gives rise to a Zeeman splitting
between up/down spin levels. We find remarkably strong
valley-dependence of 2D transport properties, and believe that the
interesting physics of valley-dependent 2D transport should be
investigated experimentally.
The recently fabricated high-mobility Si-vacuum 2D electron systems
\cite{Eng,Eng1} should be particularly suitable in this context.

We organize the rest of this article as follows: in section II 
we provide the detailed transport theory and 
a background giving a physical picture for why 2D carrier transport
should depend strongly on the valley degeneracy, followed by the
numerical results (III). In section IV we provide our calculated
mobility for the recently fabricated extreme high-mobility Si(111) 2D
samples. We 
conclude in section V with a discussion.



\section{transport theory}

To calculate the density, 
temperature, and in-plane magnetic field dependence of 2D conductivity,
$\sigma(T,n,B_{||})$, we use the Drude-Boltzmann semiclassical
theory for 2D transport limited by screened charged impurity
scattering \cite{ando,dsd_ssc}.
We assume that the 2D carrier conductivity is
entirely limited by screened impurity scattering, where the disorder
arises from randomly distributed charged impurities in a 2D plane
located at the interface between Si-SiO$_2$ (or Si-vacuum)
and random background charged impurity centers (i.e. unintentional
dopants) in the 2D layer itself.
We neglect all phonon scattering effects as well as surface
roughness scattering.
In the low temperature limit (e.g. $T<10K$ regime of interest to us)
phonon scattering is
negligible for 2D electrons in MOSFET structures and 
the short-range surface roughness scattering
at the Si-SiO$_2$ interface is only important in the high-density
limit ($n>10^{12} cm^{-2}$) (the roughness scattering can be neglected
in Si-vacuum systems).
At low carrier densities ($n < 10^{12}$ cm$^{-2}$) and at low
temperatures ($T<10K$) the 2D transport in Si-MOSFETs is dominated by
the long-range Coulomb scattering by unintentional random charged
impurities present at the Si-insulator interface 
and background unintentional dopants inside Si.
The background impurity density is low ($\sim
10^{16} cm^{-3}$), but it may dominate all other scattering in high
mobility Si-vacuum systems since the interface scattering is strongly
suppressed due to the absence of the oxide layer.

The 2D conductivity is given in the Boltzmann theory by 
\begin{equation}
\sigma=e^2 \int d\epsilon D(\epsilon)\frac{v_k^2}{2} \tau(\epsilon)\left [-
\frac{\partial f(\epsilon)}{\partial \epsilon} \right ],
\label{eq:sigma}
\end{equation}
where $D(\epsilon) = g m/(2\pi \hbar^2)$ is the density of states with
total degeneracy $g=g_s g_v$ and the carrier effective mass $m$, $v_k
= \hbar k/m$ is the carrier velocity, 
$\epsilon = (\hbar k)^2/2m$ is the usual
parabolic 2D electron energy dispersion,
$f(\epsilon)$ is the Fermi
distribution function, and $\tau(\epsilon)$ is the energy dependent transport
relaxation time. 

At $T=0$ we have $\sigma = ne^2\tau(\epsilon_F)/m$, where $n=g
k_F^2/4\pi$ is the 2D carrier density with $k_F$  being the Fermi wave
vector and $\epsilon_F$ the Fermi energy..
At finite temperatures we can express Eq.~(\ref{eq:sigma}) by
keeping the total carrier density  constant
$\sigma = {ne^2 \langle \tau \rangle}/m$,
where the energy averaged transport relaxation time $\langle \tau
\rangle$ is given by
\begin{equation}
\langle \tau \rangle = \frac {\int d\epsilon \epsilon
  \tau(\epsilon) \left [-\frac{\partial f(\epsilon)}{\partial 
  \epsilon} \right ]}  {\int d\epsilon \epsilon
   \left [-\frac{\partial f(\epsilon)}{\partial
  \epsilon} \right ]}.
\label{eq:tau}
\end{equation}
In the Born approximation the transport scattering time 
is given by considering screened
charged impurity centers 
\begin{eqnarray}
\frac{1}{\tau(\epsilon_{\bf k})} = \frac{2\pi}{\hbar} \int dz \int & &
\frac{d^2 k'}{(2\pi)^2} N_i(z) \left | \frac{U^{i}({\bf q},z)}{\varepsilon(q)}
\right |^2 \nonumber \\
&\times & (1-\cos \theta_{\bf kk'}) \delta (\epsilon_{\bf
  k}-\epsilon_{\bf k'}),
\label{eq:tau_in}
\end{eqnarray}
where $U^{i}(q;z)$ is the bare Coulomb potential for 
electron-charged impurity interaction, ${\bf q = k-k'}$ is the momentum
transfer, and 
$N_i(z)$ is the random charged impurity density in the direction ($z$)
normal to the 2D plane of confinement.
The $z=0$ is the interface plane between Si and the insulator.
In our model with two different kinds of impurity $N_i(z) = N_i^{3D} +
n_i \delta(z-z_0)$, 
where $N_i^{3D}$ is the background 3D charged impurity density and
$n_i$ is the 2D charged impurity density at $z_0$ from the interface.
In Eq.~(\ref{eq:tau_in}) $\varepsilon(q) = 1-v(q)\Pi(q)$ is the random
phase approximation (RPA) \cite{mahan} 
dielectric screening function due to the 2D electrons themselves, where
$v(q)$ is the 2D bare Coulomb interaction, and
$\Pi(q)\equiv \Pi(q,T,B_{\|})$ is the 2D finite wave
vector polarizability function depending on both temperature and
spin polarization in the presence of finite parallel magnetic field.

To understand a physical picture for why 2D carrier transport 
should depend strongly on the valley degeneracy we consider the
conductivity for a strict 2D system at the zero temperature.
However, in the realistic Si-MOSFET systems quasi-2D quantum form factor
effects arising from the finite width of
the 2D layer in $z$-direction must be included in $V^i(q;z)$ and $v(q)$.
For this purpose the usual Howard-Fang variational
function \cite{howard} is used in our numerical calculations. 
We will show that the quantum form factor effects are of considerable
quantitative importance especially at
low densities, where the system cannot really be thought of as an
almost zero-width 2D layer.

Using 2D RPA screening function at $T=0$ we have 
the scattering time for charged impurity centers
\begin{equation}
\frac{1}{\tau(\epsilon_F)} =  {2\pi\hbar} \frac{n_{i}}{m}
(\frac{2}{g})^2q_0^2 I(q_0), 
\end{equation}
where $q_0 = q_{TF}/2k_F \propto g_v^{3/2}$ ($q_{TF} =
g/a_B$ is a 2D Thomas-Fermi wave 
vector with effective Bohr radius $a_B = \hbar^2 \kappa /me^2$ where
$\kappa$ is the background dielectric constant), and
$I(q_0)$ is given by 
\begin{eqnarray}
I(q_0) & \approx & \pi + \left [ 2 + 4 \log(q_0/2) \right ]q_0 \;\;\; {\rm for}
\;\; q_0\ll1, \nonumber \\
    & \approx & \frac{\pi}{2q_0^2} \;\;\; {\rm for} \;\; q_0 \gg 1.
\end{eqnarray}
Thus in the strong screening limit (or $q_0 \gg 1$) we have
$\tau^{-1}(\epsilon_F) = \pi^2 \hbar \frac{n_i}{m} (\frac{2}{g})^2
\propto g_v^{-2}$, and the conductivity becomes $\sigma \propto
g_v^2$, i.e., the conductivity increases quadratically with $g_v$. In
the opposite limit (i.e. weak screening limit, $g_0 \ll 1$) we have
$\tau^{-1}(\epsilon_F) = 2\pi^2 \hbar \frac{n_i}{m}
(\frac{2}{g})^2q_0^2 \propto g_v$, and the conductivity becomes
$\sigma \propto g_v^{-1}$. 
In general, the effective screening ($q_0$) of
a 2D system becomes stronger as the density decreases. 
For Si(100) samples, $q_0 \approx 4.7
g^{3/2}/\sqrt{\tilde{n}}$, where $\tilde{n} = n/(10^{10}
cm^{-2})$. Thus, for the density regime of interest to us, $n <
10^{12} cm^{-2}$, $q_0 \gg 1$ and the transport of 2D Si-systems
depends strongly (quadratically) on the valley degeneracy.

At finite temperatures ($T<T_F$, where $T_F = \epsilon_F/k_B$ is the
Fermi temperature), the leading order correction to the
conductivity is linear in temperature and given by \cite{dsd_ssc}
\begin{equation}
\sigma(T) \sim \sigma_0 \left[1 - \frac{2q_0}{1+q_0}\frac{T}{T_F}
  \right ],
\end{equation}
where $\sigma_0 = \sigma(T=0)$.
In the strong screening limit ($q_0 \gg 1$) we have
\begin{equation}
\delta \sigma/\sigma_0 \propto - g_v T/n,
\label{eq:sigma_high}
\end{equation}
where $\delta \sigma = \sigma(T) - \sigma_0$. 
On the other hand, in the weak screening limit we have
\begin{equation}
\delta \sigma/\sigma_0 \propto -g_v^{5/2} T/n^{3/2}.
\end{equation}
Thus for $q_0 \gg 1$ the conductivity
decreases linearly with valley degeneracy for fixed temperature and
density, but for $q_0 \ll 1$ it decreases as $g_v^{5/2}$. 

Since there are two different carrier components (spin up and down) in the
presence of a finite 
parallel magnetic field the total conductivity of the
partially polarized system is given by 
$\sigma = \sigma_+ + \sigma_-$,
where $\sigma_{\pm} = n_{\pm}e^2\tau_{\pm}/m$ is the conductivity of
spin up ($+$) and down ($-$), respectively. $n_{\pm}$
is the carrier densities of spin state $\pm$, and
$\tau_{\pm}$ is the transport relaxation time of the spin up
(down) state. To calculate
the conductivity with screened charged impurities in the presence of
parallel magnetic field,  
spin-polarization effects must be included in the 
polarizability \cite{parB,parB1}. When the parallel magnetic field is
applied to a 2D 
electron system the polarizability becomes 
$\Pi_{tot}(q) = \Pi_{+}(q) + \Pi_-(q)$,
where $\Pi_{\pm}(q)$ is the polarizability of the spin up (down) state
and is given by at $T=0$ 
\begin{equation}
\Pi_{\pm}(q) = D_F \left [ 1  -
  \sqrt{1-\left (2k_F^{\pm}/q \right )^2} \theta (q-2k_F^{\pm}) \right ]
\end{equation}
where $D_F = {g_{\nu}m}/{2\pi}$, $g_{\nu}$ is the valley degeneracy
factor, and $k_F^{\pm}$ are 
the Fermi wave vector of the spin up (down) state. 

\begin{figure}[t]
\includegraphics[width=.90\columnwidth]{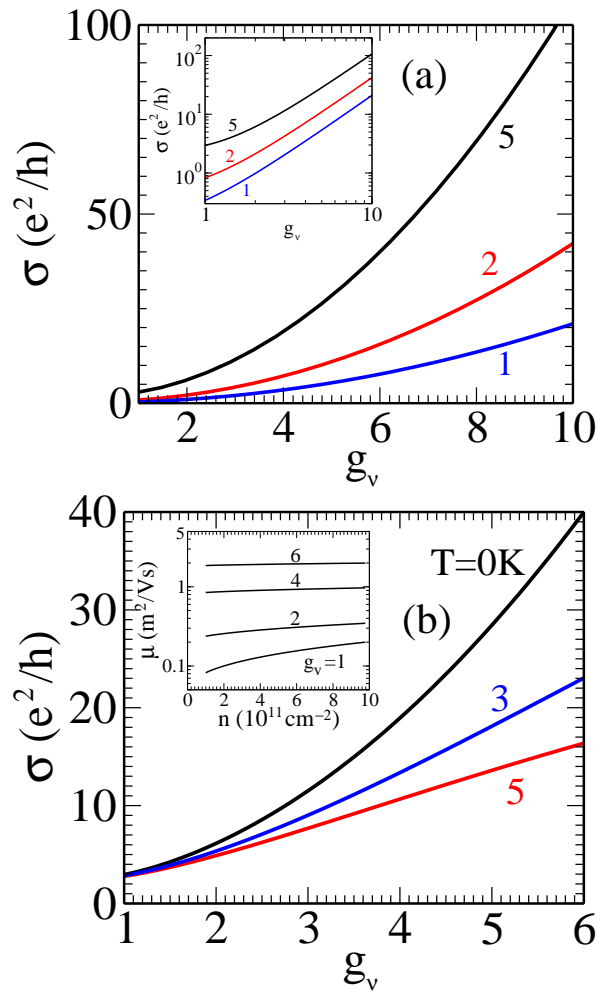}
\caption{Calculated conductivity for SiO$_2$-Si(100) MOSFET system as a function
  of valley degeneracy ($g_v$) (a) for various electron densities
  $n=$1.0, 2.0, 5.0$\times 10^{11}$ cm$^{-2}$ (bottom to
  top) at $T=0$ and (b) for a density $n=$5.0$\times10^{11}
cm^{-2}$ and for various temperatures $T=0$, 3, and 5K. 
Inset in (a) shows the same results of (a) in logarithm
  scale showing $\sigma \sim g_v^2$ for large $g_v$. 
Inset in (b) shows mobility at $T=0K$ as a function of density for different
valley degeneracies, $g_v=1$, 2, 4, 6.
The parameters corresponding 
  to Si(100) are used and the impurity density of
  $n_i=3\times10^{11}cm^{-2}$ located at the interface ($z=0$) is used. 
\label{fig:Fig1}
}
\end{figure}

For strictly 2D systems with zero thickness the spin polarization
changes the 
screening function and this effect gives rise to positive (negative)
magnetoresistance in the strong (weak) screening limits
\cite{parB,parB1}. In the strong 
screening limit 
($q_{0} \gg 1$) $\sigma(B_s)/\sigma(0) \approx 1/4$ and in the weak screening
limit ($q_{0} \ll 1$) 
$\sigma(B_s)/\sigma(0) \sim 2$, where $\sigma(0) = \sigma(B=0)$ and
$B_s$ is the magnetic field 
for complete spin-polarization. 
Since the valley degeneracy affects the screening
strength $q_0$ it is expected that the
positive magnetoresistance is enhanced as $g_v$ increases.
If we include the finite width confinement effect in the calculation
the ratio becomes much 
smaller, especially at low density. 
In addition to the screening effects in the impurity potential, for
the real systems with finite width confinement the orbital effects
\cite{parB2} 
dominate over spin effects at large magnetic fields. However we 
neglect the orbital effects in this paper, which has been considered
elsewhere \cite{parB2}.


\section{results}


\begin{figure}
\includegraphics[width=1.\columnwidth]{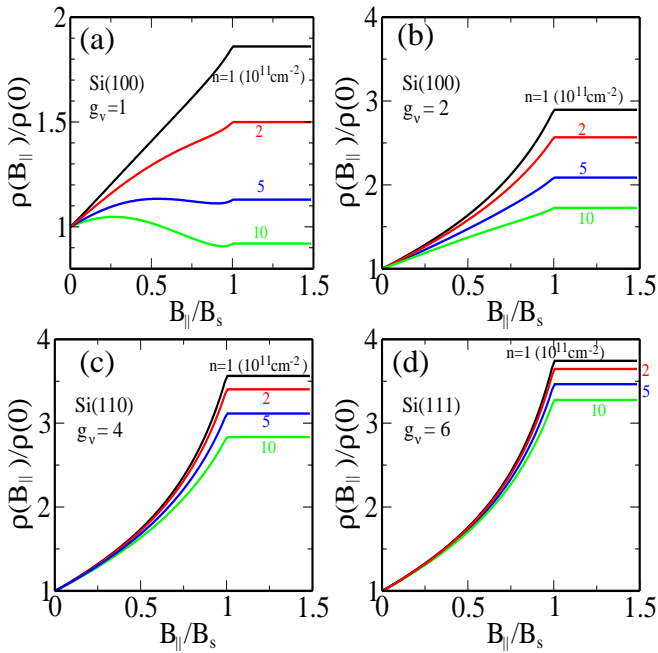}
\caption{Magnetoresistivity of SiO$_2$-Si MOSFET systems.
Magnetoresistivity ($\rho \equiv 1/\sigma$)  of (a)(b) Si(100)
($g_v=1$, $g_v=2$), (c)
  Si(110) ($g_v=4$), and (d) Si(111) ($g_v=6$) 
  MOSFET systems for various electron densities
  $n=$1.0, 2.0, 5.0, 10.0$\times10^{11} cm^{-2}$ (top to
  bottom). Impurities are located at the interface $z=0$. 
Here 
  $\rho(0)$ is the resistivity at $B=0$ and $B_s$ is the magnetic field
  for complete spin-polarization, i.e. $g_s=2$ at $B=0$ and $g_s=1$ at
  $B\geq B_s$. 
\label{fig:Fig2}
}
\end{figure}

\begin{figure}
\includegraphics[width=1.\columnwidth]{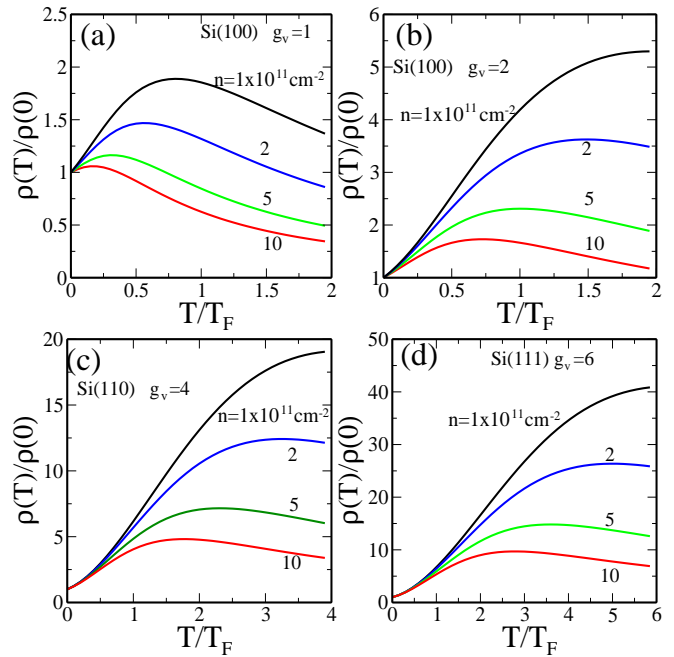}
\caption{Resistivity of SiO$_2$-Si MOSFET systems.
Temperature dependent resistivity of (a)(b) Si(100) ($g_v=1$, $g_v=2$) (c)
  Si(110) ($g_v=4$), and (d) Si(111) ($g_v=6$) 
  MOSFET systems for various electron densities
  $n=$1.0, 2.0, 5.0, 10.0$\times10^{11} cm^{-2}$ (top to
  bottom). Here 
  $\rho(0)$ is the resistivity at $T=0$ and $T_F$ is the Fermi
  temperature, $T_F=E_F/k_B$. The same parameters as Fig.~\ref{fig:Fig2}
are used.
\label{fig:Fig3}
}
\end{figure}

Throughout this paper we use the following parameters: the dielectric
constants of Si, SiO$_2$, and vacuum are $\kappa_{Si}=11.7$,
$\kappa_{SiO_2}=3.9$, and $\kappa_{vac}=1$, respectively, and the
effective masses corresponding to the Si surface of (100), (110), and
(111) are $m=0.19m_e$, $m=0.28m_e$, and $m=0.30m_e$, respectively,
where $m_e$ is the free-electron mass. In the absence of a parallel magnetic
field the spin degeneracy $g_s=2$ is used everywhere.

In Fig.~\ref{fig:Fig1}(a) we show our calculated zero magnetic field
conductivity of 
SiO$_2$-Si(100) MOSFETs as a function of valley degeneracy ($g_v$) 
for various electron densities
$n=$1.0, 2.0, 5.0$\times 10^{11}$ cm$^{-2}$ (bottom to
top) at $T=0$, by assuming that the impurities with density
  $n_i=3\times10^{11}cm^{-2}$ are located at the 
interface ($z=0$). 
As shown in the inset of Fig.~\ref{fig:Fig1}(a) in logarithm
scale, the calculated conductivity increases as
$\sigma(g_v) \propto g_v^{\alpha}$, where $\alpha$ increases as $g_v$
increases and approaches 2 which is expected in the strong screening
limit. In Fig.~\ref{fig:Fig1}(b) the conductivity is shown as a function of
$g_v$ for a density $n=$5.0$\times10^{11}
cm^{-2}$ and for various temperatures $T=0$, 3, and 5K with the same
parameters as Fig.~\ref{fig:Fig1}(a). We see that the temperature
dependence of conductivity becomes stronger since the conductivity decreases
linearly with $g_v$ for a fixed density and temperature in the strong
screening limits [see. Eq.~(\ref{eq:sigma_high})].

In Figs.~\ref{fig:Fig2} and \ref{fig:Fig3} we show the calculated
resistivity ($\rho = 1/\sigma$) for SiO$_2$-Si MOSFET systems assuming
that all impurities are located at the interface $z=0$. 
In Fig.~\ref{fig:Fig2} magnetoresistivity for (a), (b) Si(100)
with $g_v=1$, $g_v=2$, respectively, (c)
  Si(110) with $g_v=4$, and (d) Si(111) with $g_v=6$ 
  MOSFET systems are shown for various electron densities
  $n=$1.0, 2.0, 5.0, 10.0$\times10^{11} cm^{-2}$ (top to
  bottom). 
  $\rho(0)$ is the resistivity at $B=0$ and $B_s$ is the magnetic field
  for complete spin-polarization, i.e. $g_s=2$ at $B=0$ and $g_s=1$ at
  $B\geq B_s$. 
As expected the ratio $\rho(B_s)/\rho(0)$ is close to 4 for larger $g_s$
systems due to the enhancement of effective screening. Since the
finite confinement effects the ratio $\rho(B_s)/\rho(0)$ is smaller
than that of strict 2D system.
Note that the results for Si(110) and Si(111) with $g_v=2$ is very close to the
results of Fig.~\ref{fig:Fig2}(b) even though we use the parameters
corresponding to Si(110) and Si(111). The results for Si(111) with
$g_v=4$ is almost identical with the results of Fig.~\ref{fig:Fig2}(c).
The parameters corresponding to
sample properties except the valley degeneracy have only a small effect on
the ratio. The most important parameter determining the ratio is the screening
strength $q_0$. For $g_v=1$ the effective screening becomes weak, and
especially at high densities, $q_0<1$ is expected, as a consequence,
the magnetoresistance decreases as the magnetic field increases (see
the result for $g_v=1$ and  $n=10^{12} cm^{-2}$).

In Fig.~\ref{fig:Fig3} the temperature dependent resistivity of (a),(b)
Si(100) ($g_v=1$, $g_v=2$, respectively), (c) 
  Si(110) ($g_v=4$), and (d) Si(111) ($g_v=6$) 
  MOSFET systems are shown for various electron densities
  $n=$1.0, 2.0, 5.0, 10.0$\times10^{11} cm^{-2}$ (top to
  bottom). Here 
  $\rho(0)$ is the resistivity at $T=0$ and $T_F$ is the Fermi
  temperature, $T_F=E_F/k_B$. 
As shown in Eq.~(\ref{eq:sigma_high}) at low temperatures $T<T_F$ the
metallic behavior (i.e., $d\rho(T)/dT >0$) is strong for larger
$g_v$ and at small densities. Thus, a stronger metallic behavior is
expected for Si(111) with $g_v=6$. \cite{hwang} 
For $g_v=1$ effective screening is weak, i.e., 
$q_0=q_{TF}/2k_F < 1$, and strong-screening condition can only be satisfied at
very low carrier densities. Thus,
we expect rather weak temperature
and field dependence of resistivity for $g_v=1$ except at very low densities. 
Fig.~\ref{fig:Fig3} shows the very interesting feature that the crossover
temperature (i.e. the 
temperature at maximum resistivity) is almost independent of $g_v$.
Since $T_F \propto g_v/n$, the scaled crossover temperature ($T/T_F$)
increases as the valley degeneracy increases for a fixed density, but
the absolute crossover temperature is very close for all $g_s$ values.

\begin{figure}
\includegraphics[width=1.\columnwidth]{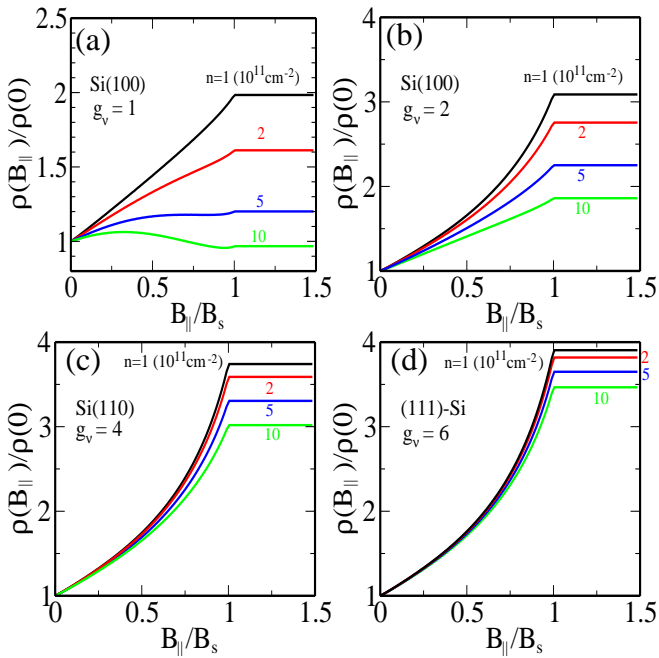}
\caption{Vacuum-Si MOSFET systems. 
Magnetoresistivity  of (a), (b) Si(100) ($g_v=1$, $g_v=2$), (c)
  Si(110) ($g_v=4$), and (d) Si(111) ($g_v=6$) 
  MOSFET systems for various electron densities
  $n=$1.0, 2.0, 5.0, 10.0$\times10^{11} cm^{-2}$ (top to
  bottom). All impurities are located at the interface $z=0$. 
\label{fig:Fig4}
}
\end{figure}

\begin{figure}
\includegraphics[width=1.\columnwidth]{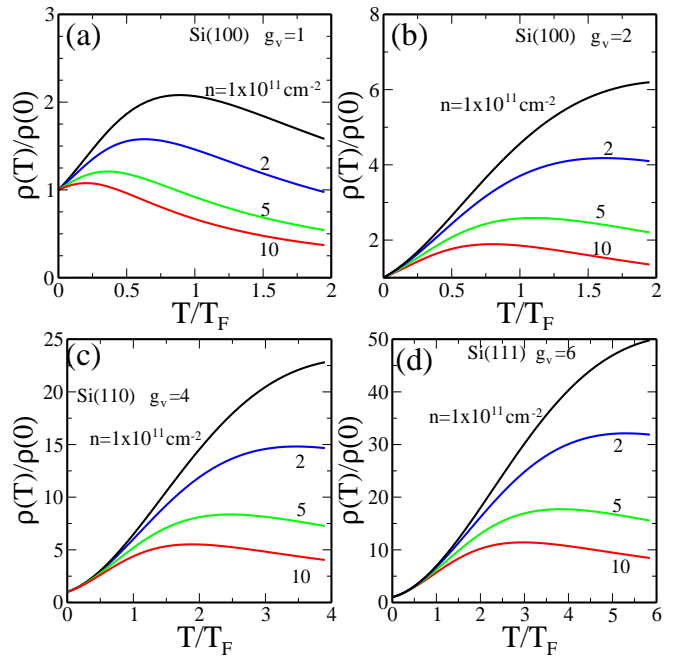}
\caption{Vacuum-Si MOSFET systems. 
Temperature dependent resistivity of (a)(b) Si(100) ($g_v=1$, $g_v=2$), (c)
Si(110) ($g_v=4$), and (d) Si(111) ($g_v=6$) 
  MOSFET systems for various electron densities
  $n=$1.0, 2.0, 5.0, 10.0$\times10^{11} cm^{-2}$ (top to
  bottom). Calculations are done with only interface impurities. 
  We use the same parameters as Fig.~\ref{fig:Fig4}.
\label{fig:Fig5}
}
\end{figure}

In Figs.~\ref{fig:Fig4} and \ref{fig:Fig5} we show the calculated
resistivity, $\rho(B_{\||})$ and $\rho(T)$, respectively, for vacuum-Si
MOSFET systems assuming 
that all impurities are located at the interface $z=0$. 
In the calculation of resistivity for vacuum-Si MOSFETs we use the same
parameters and the same impurity configuration for SiO$_2$-Si MOSFETS, 
except the insulating dielectric constant (SiO$_2$ versus vacuum,
i.e., $\kappa_{SiO_2}$ versus $\kappa_{vac}$), 
for direct comparison with the results of SiO$_2$-Si MOSFETs
(Fig.~\ref{fig:Fig2} and Fig.~\ref{fig:Fig3}). Since the screening
strength $q_0$ is inversely proportional to the background dielectric
constant at the same carrier density both magnetic field dependent and
temperature dependent resistivity for vacuum-Si MOSFETs is stronger
than for
SiO$_2$-Si MOSFETs as shown in Figs.~\ref{fig:Fig4} and
\ref{fig:Fig5}. However overall behaviors for both systems are very
similar if we assume that the impurity configurations are
identical. But in reality the impurity configurations are very different
for these two systems. In the H-passivated Si-vacuum MOSFET it is
expected that interface quality between vacuum and Si is much better
with substantially less interface charged impurities than in Si-SiO$_2$.
Especially in high mobility vacuum-Si MOSFETs it
is considered that background unintentional 3D charged impurity is the
most important scattering source.

\begin{figure}
\includegraphics[width=1.\columnwidth]{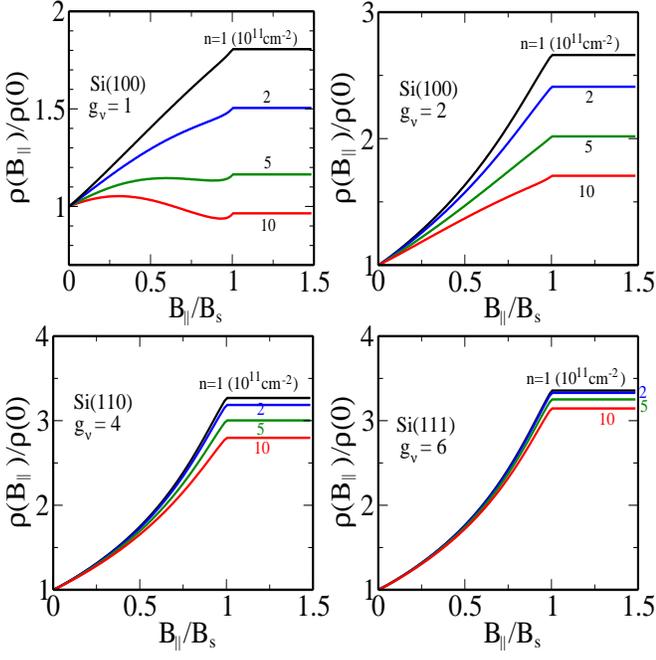}
\caption{Vacuum-Si MOSFET systems. 
Magnetoresistivity  of (a)(b) Si(100) ($g_v=1$, $g_v=2$), (c)
Si(110) ($g_v=4$), and (d) Si(111) ($g_v=6$) 
  MOSFET systems for various electron densities
  $n=$1.0, 2.0, 5.0, 10.0$\times10^{11} cm^{-2}$ (top to
  bottom). Here only unintentional 3D bulk impurities are considered with density
  $N_i^{3D}=2.3\times 10^{16}cm^{-3}$. 
\label{fig:Fig6}
}
\end{figure}

\begin{figure}
\includegraphics[width=1.\columnwidth]{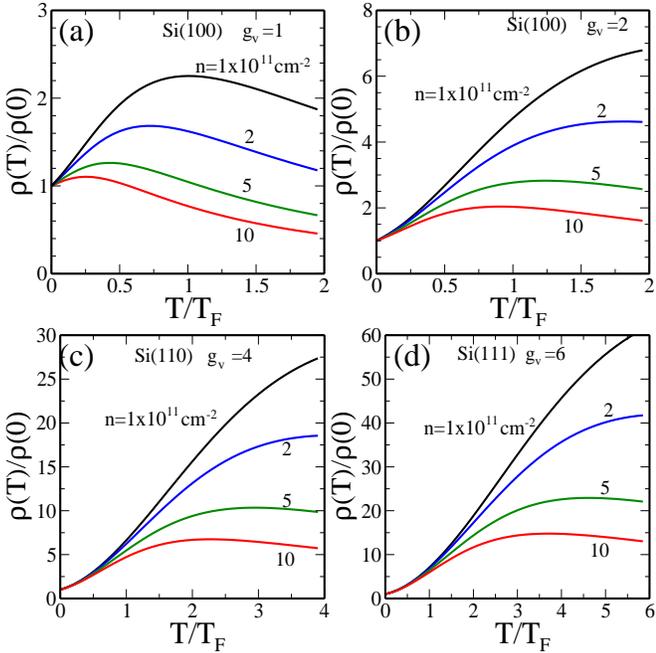}
\caption{Vacuum-Si MOSFET systems. 
Temperature dependent resistivity of (a),(b) Si(100) ($g_v=1$, $g_v=2$), (c)
Si(110) ($g_v=4$), and (d) Si(111) ($g_v=6$)
  MOSFET systems for various electron densities
  $n=$1.0, 2.0, 5.0, 10.0$\times10^{11} cm^{-2}$ (top to
  bottom). 
The same impurity configuration of Fig.~\ref{fig:Fig6} is used.
\label{fig:Fig7}
}
\end{figure}

In Figs.~\ref{fig:Fig6} and \ref{fig:Fig7} we show the calculated
magnetic field and temperature dependent
resistivity of vacuum-Si MOSFETs, $\rho(B_{\|})$ and $\rho(T)$,
respectively, assuming that 
background charged impurities with impurity density $N_i^{3D} = 2.3
\times 10^{16}cm^{-3}$ is the only  scattering source (i.e. we set the
interface impurity $n_i=0$). Even though the overall magnetic field
and temperature dependences of resistivity are very similar to the results
of Figs.~\ref{fig:Fig4} and \ref{fig:Fig5} (which are
calculated with interface charged impurity scattering)
there are substantial
differences between the two different impurity configurations. 
Comparing Fig.~\ref{fig:Fig6} with Fig.~\ref{fig:Fig4} 
we find that the 
parallel magnetic field dependence of resistivity calculated with
background 3D impurities
is weaker than that with interface impurities 
(Fig.~\ref{fig:Fig4}). However the temperature dependence of
resistivity calculated with 3D impurities is stronger than that with
interface impurity (Fig.~\ref{fig:Fig5}). 
Our predictions can be directly verified by carrying out parallel
field measurements in Si-vacuum 2D systems.

\section{High-mobility Si(111) system}
 
Very recent experimental work \cite{kane}
using ultra-clean H-passivated Si(111)-vacuum 2D
electron systems shows unprecedented high mobilities, approaching
several hundred thousand $cm^2/Vs$ at low temperatures, corresponding
to a momentum relaxation time (level broadening) of 30 ps ($\sim 0.02$
meV).
These mobility numbers of these recent samples surpass the old Si(111)
mobilities \cite{kaminsky} by factors of hundred and are
comparable in quality (in terms of the momentum relaxation time and
level broadening) to better-quality 2D GaAs electron samples where
fractional quantum Hall phenomena typically manifest.  Since the
Si(111) 2D system is a multi-valley system in contrast to the 2D GaAs 
system, the observation of the fractional quantum Hall effect in
Si(111) 2D system is a very interesting and potentially very important
new development.  
 
\begin{figure}
\includegraphics[width=1.\columnwidth]{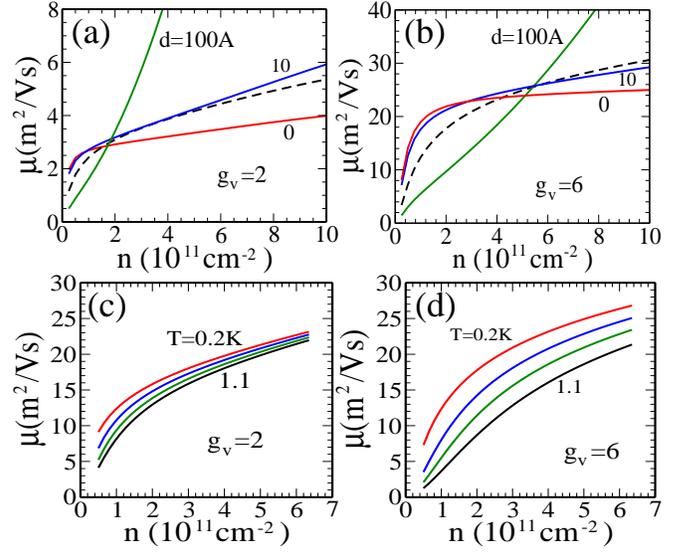}
\caption{Vacuum-Si(111) MOSFET systems. 
The calculated density dependent mobility for
  different impurity configurations at $T=0.2K$ is shown (a) for $g_v=2$ and (b)
  for $g_v=6$. The black dashed lines represent the results with only
  3D bulk impurities with $N_i^{3D} = 2.3\times 10^{16}cm^{-3}$, and
  other three lines represent results with only 2D impurities (no bulk
  impurities): $z=0A$ with $n_i=2.0\times 10^{10}cm^{-2}$, $z=10$\AA
  \; with $n_i=2.4 \times 10^{10}cm^{-2}$, and $z=100$ \AA \; with
  $n_i=21.0\times 10^{10}cm^{-2}$. (c) and (d) show the  calculated
  mobility with only 3D bulk charged impurities as a function of
  density for different temperatures $T=0.2$, 
  0.5, 0.8, and 1.1 K (from top to bottom). In (c) $g_v=2$ and the bulk 3D
  impurities of density $N_i^{3D} = 4.5\times 10^{15}cm^{-3}$ are
  used. In (d) $g_v=6$ and the bulk 3D impurities of density $N_i^{3D}
  = 2.3 \times 10^{16} cm^{-3}$ are used.
\label{fig:Fig8}
}
\end{figure}

In Fig~\ref{fig:Fig8} we show our numerical results (based on the theory given in
section II) for the low-temperature mobility (defined simply as the
conductivity divided by $ne$, $\mu = \sigma/ne$, where $n$ is the
carrier density and $e$ is 
the electron charge) of the Si(111) 2D system, both for $g_v=2$ and 6, as
functions of temperature and density for several different impurity
configurations.  The density dependence agrees very well with the
unpublished work of Kane\cite{kane}, and shows that the
disorder in this new batch of Si(111)-vacuum samples is approaching the
$\sim 10^{10} cm^{-2}$ limit which is one to two orders of magnitude lower
than the usual Si 2D samples\cite{ando}, thus explaining the
very high mobilities of these new 2D systems.  Our results also
demonstrate that the strong screening by the $g_v=6$ Si(111) system,
compared with the $g_v=2$ system, would lead to much stronger
temperature dependence of the conductivity or the mobility, thus
providing a clear means to distinguish the valley degeneracy.  Our
calculated density dependence of the mobility is approximately
consistent with the recent measurements\cite{kane}. 

Based on these calculation we can approximately estimate \cite{dean}
the expected 
activation energy of the $1/3$ fractional quantum Hall state of the
Si(111) 2D system (in these high-mobility samples) to be $\sim 5-10K$ using
the Zhang-Das Sarma model\cite{zhang} (and subtracting
out the level broadening effect \cite{dean} using our calculated
mobility).  The 
observation of the fractional quantum Hall effect in Si(111) 2D system
here is only possible because of the incredibly high mobilities
achieved through hydrogen passivation, and it is indeed a materials
science breakthrough.

\section{conclusion}

The main purpose of this article 
is a systematic theoretical investigation of the
valley degeneracy effect on 2D electronic transport properties in
Si-MOSFET systems. We calculate
theoretically charged-impurity scattering-limited 2D electronic
transport in Si(100), (110), and (111) inversion layers at low
temperatures 
and carrier densities, where screened charged impurity scattering is
important.  
The 2D mobility for a given system increases quadratically with
increasing valley degeneracy, $\mu \propto \sigma/n \propto g_v^2$,
in the strong screening limit ($q_0 = q_{TF}/2k_F$) for the same
impurity configuration. We also show that
the temperature and 
the parallel magnetic field dependence of the 2D conductivity is
strongly enhanced by increasing the valley degeneracy. 
All our results are valid only at carrier densities above which
localization effects become important, but we estimate that the
transition to the insulating state occurs well below $10^{11}$
cm$^{-2}$ density in the high mobility systems of our interest
\cite{Eng,Eng1,kane} in the current work.

We conclude by emphasizing our findings in both
Si-SiO$_2$ and Si-vacuum MOSFETs.
The parallel magnetic field and the temperature
dependence of the resistivity (at zero parallel field)
manifest strong valley dependence regardless of impurity configurations.
For $g_v=1$ both $\rho(T)$ and $\rho(B_{\|})$ 
show weak temperature and magnetic field dependence by virtue of weak
screening, and 
for $g_v=6$ due to strong screening (i.e. large $q_0 = q_{TF}/2k_F$)
the resistivity  
shows both strong temperature and magnetic field dependence.
Our finding of remarkably strong
valley-dependence of 2D transport properties in Si-MOSFETs and the
interesting physics of valley-dependent 2D transport should be
investigated experimentally.
Similar strong valley degeneracy dependence is also apparent in the
many-body effects of 2D systems, which have been studied elsewhere
\cite{li}.
We have also provided detailed calculations for the valley-dependent
transport properties of 2D Si(111) systems as a function of
temperature and density in very high-mobility low-disorder samples,
commenting on the possible activation energy for the fractional
quantum Hall effect in such ultraclean Si system.

\section*{acknoledgements}
This work was supported by LPS-NSA.

\end{document}